\documentclass[twoside,reqno]{fich}
\usepackage{epsfig,cite}
\usepackage{amssymb,amsmath}
\usepackage{times}
\begin{document}
\sloppy \raggedbottom
\setcounter{page}{1}
%%%%%%%%%%%%%%%%%%%%%%%%%%

\newcommand{\beq}{\begin{equation}}
\newcommand{\eeq}{\end{equation}}
\newcommand{\beqa}{\begin{eqnarray}}
\newcommand{\eeqa}{\end{eqnarray}}
\newcommand{\noi}{\noindent}
\newcommand{\g}{{\mathfrak g}}
\newcommand{\e}{{\varepsilon}}
\def\gr{\text{gr}}
\def\lb{\left\{\left|}
\def\rb{\right|\right\}_{_{\hskip -.05 truecm N}}}

\newpage
\setcounter{figure}{0}
\setcounter{equation}{0}
\setcounter{footnote}{0}
\setcounter{table}{0}
\setcounter{section}{0}


% use the thanks command within \title, \author or \address for
%footnotes: % \title{Title} or  \title{Title\thanks{...}}

\title{Parafermions, ternary algebras and their  associated superspace}

\runningheads{R. Campoamor-Stursberg and M. Rausch de Traubenberg}{Parafermions
and ternary algebras}

\begin{start}

% \author{Name1}{aff.label1},
%\coauthor{Name2}{aff.label2},
%\coauthor{Name3}{aff.label3}
%\address{Address1}{aff.label1}
%\address{Address2}{aff.label2}
%\address{Address3}{aff.label3}

\coauthor{R. Campoamor-Stursberg}{1},
\author{M. Rausch de Traubenberg}{2}

\address{I.M.I-U.C.M, Plaza de Ciencias 3, E-28040 Madrid, Spain}{1}
\address{ IPHC-DRS, UdS, CNRS, IN2P3; 23  rue du Loess,
Strasbourg, 67037 Cedex, France}{2}

%you may repeat \coauthor as many times as you need
%names may have more than one aff.label, e.g.,
%\coauthor{Name2}{aff.label2,aff.label3},

\begin{Abstract}
Parafermions of order two  are shown to be the
fundamental tool to construct ternary superspaces related to cubic
 extensions of the Poincar\'e algebra.
\end{Abstract}
\end{start}

%%%%%%%%%%%%%%%%%%%%%%%%%%%%%%%%%%%%%%%%%%%%%%%%%%%%%%%%%%%%%
% The main text of your paper                               %
%%%%%%%%%%%%%%%%%%%%%%%%%%%%%%%%%%%%%%%%%%%%%%%%%%%%%%%%%%%%%

\section{Introduction}
Ternary (and in general $n-$ary) algebras have been considered for
some time in the literature from the purely mathematical point of
view, and only recently there has been some revival of these
structures in connection with physics, like in the field theory of
multiple $M2$-branes, where certain metric $3$-algebras appear
naturally \cite{blg} (see also \cite{mfm} for a formal
description).  In a different context it has been observed that
some cubic extensions of the Poincar\'e algebra can be implemented
into the Quantum Field Theory frame \cite{cubic} (the underlying
algebraic structure, called Lie algebras of order three, having
been introduced in \cite{flie}).

It should however be taken into account that ternary (and higher
order ones) structures are quite different from their quadratic
analogue, the Lie algebras and superalgebras. In particular, for
the extensions considered in \cite{flie}, the various cubic
brackets (see below) do not allow us to order a given monomial in
a definite order. This means specifically that finite dimensional
representations are automatically non-faithful (see the second
paper of \cite{flie} or \cite{cours}). The latter obstruction is
certainly only one among the various reasons that justify the
formal difficulties encountered in order to construct an
appropriate ternary superspace. However, despite these
difficulties, it has recently been realised that Lie algebras of
order three share some similarities with Lie superalgebras.
Indeed, a formal study of Lie algebras of order three enables us
to identify the parameters of the corresponding transformation and
then to define groups associated to Lie algebras of order three.
It turns out that the fundamental variables which naturally
describe the parameters of the transformation correspond to the
genuine cubic extension of the Grassmann algebra, called the
three-exterior algebra (see \eqref{roby} below) introduced by Roby
\cite{roby}. These similarities allow the construction of linear
representations of groups associated to Lie algebras of order
three, in terms of matrices, the entries of which belong to the
three-exterior (or Roby) algebra \cite{hopf}, in straight analogy
with Lie supergroups.

It is then natural to construct a ternary superspace using the
Roby variables. In this paper, we give a step-by-step construction
of the ternary superspace associated to the  cubic extension of
the Poincar\'e algebra \eqref{3poin} (see below). At the very end,
the fundamental variables needed to define a ternary superspace
turn out to be the order two parafermions. Parafermions and more
generally parastatistics were introduced a long time ago as an
exotic possibility extending the Bose and Fermi statistics
\cite{green,para}. In particular, order two parafermions satisfy
cubic relations, the latter allowing us to generate a ternary
algebra. It is very interesting to notice that two different
structures, which have {\it a priori} no relation, can be unified
by this ansatz. The question whether these two structures
(parafermions and Lie algebras of order three) have some further
hidden relations arises at once.

Let us we mention that parafermions and parabosons have also been
considered  in a rather different context \cite{ter,parag,ply}. It
is also important to mention that ternary superspaces were defined
by several authors in one or two space-time dimensions, where the
situation is somewhat exceptional (the Lorentz algebra being
either trivial or abelian) \cite{super, az}.

The contents of this paper is the following. In section 2, the
basic definitions of Lie algebras of order three, together with
the specific cubic extension of the Poincar\'e algebra relevant
for the sequel are given. Section 3 is devoted to the explicit
construction of the ternary superspace. Finally, some conclusions
and perspectives are given in Section 4.

\section{Lie algebra of order three and cubic extensions of the
Poincar\'e algebra}

In this section, we recall the basic properties of Lie algebras
of order three. We also recall how a
 cubic extension of the Poincar\'e
algebra  arises in this context. Higher order algebraic structures
(in fact $F-$ary), called Lie algebras of order $F$ and
generalising Lie (super)algebras, were introduced in \cite{flie}.
In this note, we  are mostly interested in elementary  real Lie
algebras of order three. An elementary (real) Lie algebra of order
three is given by $\g= \g_0 \oplus \g_1 = \left<X_i,  \
i=1,\cdots, \dim \g_0\right> \oplus \left<Y_a, \ a=1,\cdots, \dim
\g_1\right>$ where $\g_0$ is a real Lie algebra and $\g_1$ is a
real representation of $\g_0$, satisfying the following brackets

\beqa
\label{flie}
\left[X_i,X_j\right]&=& f_{ij}{}^k X_k,\nonumber  \\
\left[X_i,Y_a\right]&=& R_{ia}{}^b Y_b, \nonumber \\
\left\{Y_a,Y_b,Y_c\right\}&=&
Y_a Y_b Y_c + Y_b Y_c Y_a + Y_c Y_a Y_b  \\
& +& Y_a Y_c Y_b + Y_b Y_a Y_c +
 Y_c Y_b Y_a = Q_{abc}{}^iX_i, \nonumber
\eeqa

\noi and fulfilling the following fundamental identities for any
$Y_a, Y_b, Y_c, Y_d$ in $\g_1$

{\small
\beqa
\label{Jac}
0&=&[Y_a,\left\{Y_b,Y_c,Y_d\right\}]+
[Y_c,\left\{Y_d,Y_a,Y_b\right\}]\nonumber \\
&+&[Y_b,\left\{Y_c,Y_d,Y_a\right\}] +
[Y_d,\left\{Y_a,Y_b,Y_c\right\}].
\eeqa
}

\noi Looking at the various brackets, one immediately observes
that a Lie algebra of order three is endowed with two different
products: one binary given by the usual commutator, and a ternary
given by a fully symmetric product. Furthermore, a direct
inspection of \eqref{flie} and \eqref{Jac} shows that Lie algebras
of order three are a ternary extension of Lie superalgebras, where
the anticommutator is replaced by a fully symmetric cubic bracket.
Moreover, the second equation of \eqref{flie} is just a
consequence of our assumption that $\g_1$ is a representation of
$\g_0$ which is specified by the matrices $R_i$. This
representation is denoted with ${\cal D}$. Finally, the last
equations just assume that $\g_0 \subseteq {\cal S}^3(\g_1)$
(where ${\cal S}^3(\g_1)$ is the three-fold symmetric tensor
product of $\g_1$). Many examples of Lie algebras of order $F$
were given in \cite{flie}, and a formal study of this algebraic
structures was initiated \cite{gr, hopf, color}.

Having introduced the ternary algebra \eqref{flie}, one immediately
may wonder if it could be applied in physics. In fact,
among various possibilities, the cubic extension of the Poincar\'e
algebra $I\mathfrak{so}_3(1,3)= \g_0 \oplus \g_1$, with $\g_0=
\left<L_{\mu \nu} = -L_{\nu \mu}, \ P_\mu, 0 \le \mu, \nu \le 3
\right>$ generating the Poincar\'e algebra and $\g_1= \left<
V_\mu, 0 \le \mu \le 3 \right>$ being the vector representation,
and brackets

\begin{eqnarray}  \label{3poin}
\left[L_{\mu \nu }, L_{\rho \sigma}\right]&=& \eta_{\nu \sigma } L_{\rho \mu
}-\eta_{\mu \sigma} L_{\rho \nu} + \eta_{\nu \rho}L_{\mu \sigma} -\eta_{\mu
\rho} L_{\nu \sigma},  \notag \\
\left[L_{\mu \nu }, P_\rho \right]&=& \eta_{\nu \rho } P_\mu -\eta_{\mu \rho
} P_\nu, \nonumber \\
 \left[L_{\mu \nu }, V_\rho \right]&=& \eta_{\nu \rho } V_\mu
-\eta_{\mu \rho } V_\nu, \  \\
 \left[P_{\mu}, V_\nu \right]&= &0, \nonumber \\
 \{ V_\mu, V_\nu, V_\rho \}&=& \eta_{\mu \nu } P_\rho + \eta_{\mu \rho }
P_\nu + \eta_{\rho \nu } P_\mu,  \notag
\end{eqnarray}

\noi where $\eta_{\mu \nu} = \text{diag}(1,-1,-1,-1)$ is the
Minkowski metric was intensively studied in the framework of
Quantum Field Theory \cite{cubic, noether,cours}. This ternary
extension of the Poincar\'e algebra is non-trivial in the sense
that a space-time translation is generated by a cubic composition
of three elements of $\g_1$. Since, $\g_1$ is in the vector
representation of the Lorentz algebra, differently from
supersymmetry, a multiplet contains states of different spin, but
underlying the same statistics \cite{cubic}.

\section{Ternary superspace}
The representation theory of \eqref{3poin} was analysed in
\cite{cubic}, and invariant Lagrangians were explicitly
constructed. However, since our basic algebra is cubic instead of
quadratic, the construction of a ternary superspace is more
involved. For instance, looking the the fifth relation in
\eqref{3poin}, one observes that we cannot order a monomial in $V$
in a definite order. And, in particular, defining the universal
enveloping algebra ${\cal U}(\g)$ a Poincar\'e-Birkhoff-Witt was
established, and it was shown \cite{hopf} that ${\cal U}(\g_1)$ is
isomorphic (as a vector space) to the three-exterior algebra (see
\eqref{roby}), which in turn is infinitely generated \cite{roby}.
 This means  that if we consider a finite dimensional representation of
 \eqref{3poin}, it cannot be faithful (see {\it e.g.} \cite{flie,cours}). This observation
is certainly one of the reasons for the difficulty to construct a
ternary superspace. In this section we show one possibility of
associating an appropriate ternary superspace to the algebra
\eqref{3poin}. In the following construction, the various
assumptions leading to the ternary superspace are introduced
step-by-step. The result of this section was given in \cite{tern}.

\subsection{The natural variables}
Starting from a Lie algebra of order three $\g = \g_0 \oplus
\g_1=\left<X_i,  \ i=1,\cdots, \dim \g_0\right> \oplus \left<Y_a,
\ a=1,\cdots, \dim \g_1\right>$, and using the Hopf algebra
techniques,  it can be shown that \cite{hopf}

\begin{enumerate}
\item to each generator  of $\g_0$, one associates a commuting parameter
$X_i \to \alpha^i$;
\item to each generator of $\g_1$, one associates a variable $Y_a \to \eta^a$,
such that the variables $\eta^1,\cdots, \eta^{\dim \g_1}$
generate the three exterior algebra

\beqa
\label{roby}
\eta^a \eta^b \eta^c +
\eta^b \eta^c \eta^a +
\eta^c \eta^a \eta^b +
\eta^a \eta^c \eta^b +
\eta^b \eta^a \eta^c +
\eta^c \eta^b \eta^a =0.
\eeqa

\noi Furthermore, the variables $\eta^a$ are in the dual
representation of ${\cal D}$. It has to be mentioned that the
algebra generated by the $\eta$'s can either be  real or complex.
From now on, we are considering only real Roby algebras.
\end{enumerate}

It is then natural in the case of the cubic extension \eqref{3poin}, to
postulate that the ternary superspace is generated by

\beqa
\label{super}
X=(x^\mu,\theta^\mu),
\eeqa

\noi
where $x^\mu$ are the space-time coordinates and $\theta^\mu$ are their
ternary analogues which are in the vector representation of the
Lorentz group and which satisfy the algebra  \eqref{roby}.

\subsection{Realisation of the Poincar\'e algebra}

The next step in our construction is to define differential
operators which act on the ternary superspace \eqref{super} and
which realise the Poincar\'e algebra. It is then necessary to
introduce some variables conjugate the the variables $X$. Denote
$P_\mu$ (resp. $\partial_\mu$) the conjugate variables of $x^\mu$
(resp. $\theta^\mu$). The action of $P_\mu$ on $x^\mu$ is
straightforward ($[P_\mu,x^\nu]=\delta_\mu{}^\nu$). Following
Green \cite{green,para}, the more general quantisation which
ensures that $\theta^\mu$ are vectors of the Lorentz algebra is
given by the parafermions \footnote{Or parabosons, but the
parabosonic algebra turns out to be incompatible with requirement
\eqref{roby}.}. We thus assume the parafermionic relations

\beqa \label{para1}
\begin{array}{ll}
\left[\left[\theta^\mu,\theta^\nu\right],\theta^\rho\right]=0, &
\left[\left[\theta^\mu,\theta^\nu\right],\partial_\rho\right]=
-\delta^\mu{}_\rho \theta^\nu+\delta^\nu{}_\rho \theta^\mu, \\
\left[\left[\theta^\mu,\partial_\nu\right],\theta^\rho\right]=
\delta_\nu{}^\rho \theta^\mu, &
\left[\left[\theta^\mu,\partial_\nu\right],\partial_\rho\right]=
-\delta^\mu{}_\rho \partial_\nu,\\
\left[\left[\partial_\mu,\partial_\nu\right],\theta^\rho\right]=
-\delta_\mu{}^\rho \partial_\nu+\delta_\nu{}^\rho \partial_\mu
 &
\left[\left[\partial_\mu,\partial_\nu\right],\partial_\rho\right]=0.
\end{array}
\eeqa

\noi As a consequence, if we define \beqa \label{para4} {\cal
J}_{\mu \nu}= [\theta_\nu, \partial_\mu]-
[\theta_\mu,\partial_\nu], \eeqa

\noi the relations \eqref{para1} ensure that \eqref{para4} act
correctly on $\theta$ and $\partial$: \beqa [{\cal J}_{\mu \nu},
\theta_\rho]= \eta_{\nu
\rho} \theta_\mu - \eta_{\mu \rho} \theta_\nu. \eeqa

\noi  This means that  $P_\mu$ and

\beqa \label{lorentz} L_{\mu \nu}
= x_\nu P_\mu -x_\mu P_\nu +{\cal J}_{\mu \nu}, \eeqa

\noi
generate the Poincar\'e transformations upon the ternary
superspace \eqref{super}.

\subsection{Order two parafermions}

Since we are considering ternary algebras involving fully
symmetric products, putting \eqref{roby} and \eqref{para1}
together shows explicitly that we are considering parafermions of
order two. This means that the relations \eqref{roby} have to be
supplemented by the additional conditions \cite{para}:

 \beqa \label{para}
\left\{\theta^\mu,\theta^\nu,\theta^\rho\right\}&=&0, \nonumber \\
\left\{\theta^\mu,\theta^\nu,\partial_\rho\right\} &=&2
\delta^\mu{}_\rho \theta^\nu+
2 \delta^\nu{}_\rho \theta^\mu, \nonumber \\
\left\{\theta^\mu,\partial_\nu,\partial_\rho\right\}&=& 2
\delta^\mu{}_\nu \partial_\rho+
2 \delta^\mu{}_\rho \partial_\nu, \\
\left\{\partial_\mu,\partial_\nu,\partial_\rho\right\}&=&0.
\nonumber \eeqa

\noi It is interesting to observe that the construction leading to
\eqref{para} and \eqref{para1} goes in reverse order to that of
parafermionic algebras. Historically, parafermions were defined by
means of equation \eqref{para1}, in order to realise the Lorentz
algebra. After all the order of paraquantisation (here two, but
in general $p$) is specified by assuming on  which representation
of the Lorentz algebra the parafermionic algebra acts. Order $p$
parafermionic algebras involved fully symmetric brackets of order
$p+1$ and, in particular, order two parafermionic algebra give
rise to the brackets \eqref{para}. However, in our construction,
the cubic brackets \eqref{roby} are obtained from the very
beginning, by our superspace assumption. Finally, notice that the
order two parafermionic algebra  \eqref{para1}, \eqref{para} is a
non-faithful representation of the algebra \eqref{roby}
since with respect to the Roby algebra
we have one more relation $[[\theta^\mu,\theta^\nu],\theta^\rho]=0$.
(In particular the Roby algebra is infinitely many generated
\cite{roby, hopf}
although the order three parafermionic algebra is finite dimensional).

As a final observation of this section, let us mention that if we
define $\psi_\pm{}_\mu=\theta_\mu \pm \partial_\mu$ we have

\beqa
\label{3poin*}
\{\psi_\pm{}_\mu,\psi_\pm{}_\nu,\psi_\pm{}_\rho\}=
\mp 4(\eta_{\mu \nu} \psi_\pm{}_\rho+
\eta_{\nu \rho} \psi_\pm{}_\mu+
\eta_{\rho \mu} \psi_\pm{}_\nu)
\eeqa

\noi
which is very similar to the fifth equation of \eqref{3poin}.
This is certainly not a  coincidence.
Indeed, as we have seen ${\cal U}(\g)$ can be endowed with
a Hopf algebra structure \cite{hopf}, and in particular with a coproduct.
This is precisely this coproduct, that makes
that three-exterior algebra \eqref{roby} emerges naturally
on ${\cal U}(\g_1)^*$ (the dual of  ${\cal U}(\g_1)$). The introduction
of the conjugate momenta of $\theta^\mu$ lead to \eqref{para}
and finally to \eqref{3poin*}.

\subsection{Realisation of the ternary part of the algebra}

Now, we would like to construct a differential realisation of the
purely ternary part of the algebra. The relations  \eqref{para1}
shows that the natural relations upon the $\theta$'s and the
$\partial$'s involve double commutators. This means in particular
that we cannot expect to
 construct a differential operator
$V_\mu$  from  $\partial_\mu$ and $\theta^\mu$ acting on
$\theta^\mu$ and satisfying the cubic relations \eqref{3poin}. For
instance, if we assume that $V_\mu = \partial_\mu + \cdots$, we
obtain that $[V_\mu,\theta^\nu]= [\partial_\mu,\theta^\nu]
+\cdots$. But since  the relations \eqref{para1} and \eqref{para}
are cubic, there is no bilinear relations upon $\theta^\mu$ and
$\partial_\nu$ and consequently $[\partial_\mu,\theta^\nu]$
emerges as a new object.

 This situation is very similar to
the implementation of the Noether theorem within the framework of
ternary symmetries, where the conserved charges generate the
symmetry through quadratic relations using the usual quantisation
procedure ({\it e.g.} the equal-time (anti)-commutation relations).
Let us briefly recall how it works. In \cite{cubic} we have
obtained some representations of the algebra \eqref{3poin}
$P_\mu \to \tilde P_\mu, L_{\mu \nu } \to \tilde L_{\mu \nu}$
and $V_\mu \to \tilde V_\mu$ acting on a  multiplet $\Phi$
(in fact various multiplets were obtained). Next,
we have constructed an invariant Lagrangian ${\cal L}(\Phi)$
and obtained the associated conserved charges \cite{cubic, noether}
 $\hat L_{\mu \nu}, \hat P_\mu, \hat V_\mu$. For instance, $\hat V_\mu$
is given by

$$
\hat V_\mu = -i \int d^4 x \frac{\partial {\cal L }}{\partial \partial_0 \Phi}
\tilde V_\mu \Phi
$$

\noi
(standard expressions  were obtained for $\hat P_\mu$ and $\hat L_{\mu \nu}$),
and is such that after quantisation

\beqa
\label{not}
\tilde V_\mu \Phi=[\hat V_\mu,\Phi]
\eeqa
\noi
(in the usual way the Poincar\'e transformations of
the multiplet $\Phi$ are given by $ [\hat L_{\mu \nu}, \Phi]
,[\hat P_\mu,\Phi]$).
It is important to emphasise that at this point we are dealing with
{\it usual} bosonic and fermionic fields satisfying the standard
(anti-)com\-mutation relations {\it i.e.} there is no need to introduce
some fields with exotic behaviour in order to obtain \eqref{not}.
Next, we have shown that the algebra is realised through
multiple-commutators

\beqa \label{adj} [\hat V_\mu,[\hat V_\nu, [\hat V_\rho,\Phi]]] +
\text{perm.}=
 \eta_{\mu \nu} [\hat P_\rho, \Phi]+
\eta_{\nu \rho} [\hat P_\mu,\Phi] + \eta_{\mu \rho}  [\hat
P_\nu,\Phi]. \eeqa

\noi This procedure is standard in the implementation of Lie
(super)algebra in Quantum Field Theory, but the equation
corresponding to \eqref{adj} in this case is not the end of the
story  since the Jacobi
identities allow to obtain a relation which is independent of the
fields $\Phi$. But here, in the context of ternary symmetries, the
situation is very different, since the fundamental identities
\eqref{Jac} do not allow to write the algebra in a $\Phi$
independent form. This weaker realisation of the algebra has the
interesting consequence that it enables us to  consider algebraic
structure (in Quantum Field Theories), different from Lie
superalgebras,  without contradicting  the spin-statistics theorem
(see \cite{cours} for a discussion). Finally, it is a matter of
calculation to check that the fundamental identities \eqref{Jac}
 are satisfied by
the realisation \eqref{adj}.

We thus see that the implementation of  Noether theorem
in ternary algebras presents
some similarities with the natural action
defined on parafermions.
Indeed, in both cases the natural objects are
the commutators (or the anticomutators for fermions). 
This suggests to try to realise \eqref{3poin}
on the superspace \eqref{super} in the form of \eqref{adj}.
We introduce  the
parameters of the transformations $\e^\mu$ (of the
same nature of $\theta$ and as such satisfying \eqref{roby}, \eqref{para}
and \eqref{para1})
such that we have the
transformation
$$
\theta^\mu \to \theta'{}^\mu=\theta^\mu + \e^\mu,
$$
\noi under \eqref{3poin}.
  Now  we define the generator

 \beqa \label{rep}
V=\left[\e^\mu,\partial_\mu\right]+ \left[\theta,\theta^\mu\right]
\left[\e^\sigma, \theta_\mu\right]P_\sigma, \eeqa

\noi
such that

\beqa \label{trans} \delta
\theta^\alpha=\left[V,\theta^\alpha\right]=\e^\alpha, \ \ \delta
x^\alpha=\left[V,x^\alpha\right]= \left[\theta,\theta^\mu\right]
\left[\e^\alpha, \theta_\mu\right]. \eeqa

\noi It is important to realize that  the $\delta x^\alpha$'s are
commuting real variables. Two observations are in order here.
Firstly, in the realisation of the algebra, due to the nature of
the para-commutation relations \eqref{para1}, it is not possible
to dissociate the generators $V_\mu$ and the parameters $\e^\mu$.
Secondly, in order to have the appropriate transformations
properties for $\theta^\mu$, we are forced to introduce one more
parafermionic variable $\theta$ in the scalar representation of
the Lorentz algebra. This new variable can be seen, together with
$\theta^\mu$, to be some parafermion associated to
$\mathfrak{so}(1,4)$.

\subsection{Closure of the algebra}
Now we have to check the closure of the algebra in the form
\eqref{adj}. In particular, if we compute

\beqa \label{leib}
[V_1,[V_2,[V_3,\theta^{\alpha_1}\theta^{\alpha_2}
\theta^{\alpha_3}]]]&=& \e_1^{\alpha_1}  \e_2^{\alpha_2}
\e_3^{\alpha_3} + \e_2^{\alpha_1}  \e_3^{\alpha_2} \e_1^{\alpha_3}
+ \e_3^{\alpha_1}  \e_1^{\alpha_2}   \e_2^{\alpha_3}\nonumber \\ &
+& \e_1^{\alpha_1}  \e_3^{\alpha_2}   \e_2^{\alpha_3} +
\e_2^{\alpha_1}  \e_1^{\alpha_2}   \e_3^{\alpha_3} +
\e_3^{\alpha_1}  \e_2^{\alpha_2}   \e_1^{\alpha_3}, \eeqa

\noi we observe that it is fully symmetric with respect to the
indices $1,2,3$. This means that $[V_1,[ V_2,[
V_3,\theta^{\alpha_1}\theta^{\alpha_2} \theta^{\alpha_3}] +
\text{perm}$.  never vanishes. From now, in order to simplify the
notations, we denote $V_1.V_2.V_3.
\theta=[V_1,[V_2,[V_2,\theta]]]$ and $\{V_1,V_2,V_3\}. \theta =
[V_1,[V_2,[V_2,\theta]]] + \text{perm.}$, {\it etc.}

If one proceeds along these lines in the case of supersymmetry,
ones obtains the same kind of results. However, in this case the
closure of the algebra is guaranteed by the introduction of the
Grassmann (anti-commuting) variables and consequently,  the
anti-commutators get replaced by the commutators. There is some
analogous substitution in the context of ternary algebras but here
the situation is more involved because the variables $\e^\mu$ do
not satisfy quadratic relations.

Indeed, we have shown in \cite{hopf} that ternary algebras of
order three inherit of a $\mathbb Z_3-$graded structure, or more
precisely of a $\mathbb Z_3-$twisted tensor product. This means in
particular, that if we consider three successive transformations
specified by $\e_1,\e_2,\e_3$ we get a $\mathbb Z_3 \times \mathbb
Z_3 \times \mathbb Z_3-$graded structure. This $\mathbb
Z_3-$structure implies that, taking the parameters of the
transformation, the bracket of order three is no longer fully
symmetric, but as to be defined with the cubic primitive root of
unity that we denote by $q$. This is a kind of Jordan-Wigner
transformation adapted to ternary algebras. In fact these types of
structures, where the brackets are neither symmetric nor
antisymmetric have been considered before in the literature, even
for quadratic algebras (as a possible generalisation of Lie
(super)algebras) and have been called colour algebras
\cite{coloralg}. The basic tool to define colour Lie
(super)algebras is a grading determined by an Abelian group. The
latter, besides defining the underlying grading in the structure,
moreover provides a new object known as commutation factor
associated to an Abelian group $\Gamma$ (here $\Gamma= \mathbb Z_3
\times \mathbb Z_3 \times \mathbb Z_3$).
 A commutation factor $N$ is a
map $N: \ \Gamma \times \Gamma \to \mathbb
C\setminus\left\{0\right\}$
 satisfying the
following constraints:

\begin{enumerate}
\item  $N(a,b) N(b,a)=1, $ for all $a,b \in \Gamma$; \item
$N(a,b+c)= N(a,b) N(a,c),$ for all  $a,b,c  \in \Gamma$; \item
$N(a+b,c)= N(a,c) N(b,c),$ for all $a,b,c  \in \Gamma$.
\end{enumerate}

\noi
In our case, the commutation factor is given by

\beqa
\label{3-com}
N(\vec a, \vec b)= q^{a_1(b_2 + b_3) + a_2 b_3 -
 b_1(a_2 + a_3) -b_2 a_3},
 \eeqa

\noi where $q=e^{\frac{2i \pi}{3}}$ and $\vec a, \vec b \in
\mathbb Z_3^3$. Then in the same vain of the colour algebras,
colour algebras of order three may be defined \cite{color}. In
particular, defining

\beqa
\label{3-color}
\lb V_1, V_2, V_3 \rb &=& V_1 V_2 V_3 +
N\Big(\gr(\e_1),\gr(\e_2)+\gr(\e_3)\Big) V_2 V_3 V_1
  \nonumber \\
&+&
 N\Big(\gr(\e_1)+\gr(\e_2),\gr(\e_3)\Big)  V_3 V_1 V_2   \\
&+& N\Big(\gr(\e_2),\gr(\e_3)\Big) V_1 V_3 V_2+
 N\Big(\gr(\e_1),\gr(\e_2)\Big)V_2 V_1 V_3 \nonumber \\
&& +N\Big(\gr(\e_1),\gr(\e_2)\Big) N\Big(\gr(\e_1),\gr(\e_3)\Big)
\times \nonumber\\
&\times&
N\Big(\gr(\e_2),\gr(\e_3)\Big)V_3 V_2 V_1, \nonumber \eeqa

\noi with
$\text{gr}(\e_1)=(1,0,0),\;
\text{gr}(\e_2)=(0,1,0),\; \text{gr}(\e_3)=(0,0,1)$,
the cubic
brackets \eqref{3-color} adopt the following form
(there is also  corresponding fundamental identities, but there
are not relevant for our purpose \cite{color})

\beqa
\label{col3}
\lb V_1, V_2, V_3 \rb
&=&V_1 V_2 V_3 +  q^2 V_2 V_3 V_1 + q^2 V_3 V_1 V_2 \\
&+& q V_1 V_3
V_2 + q V_2 V_1 V_3 + V_3 V_2 V_1.
\nonumber
\eeqa

\noi In particular, since the constraint $1+q+q^2=0$ is satisfied
and $V_1 .V_2 .V_3.(\theta^{\alpha_1} \cdots \theta^{\alpha_n})$
is fully symmetric in the subindices $1,2,3$, we automatically
have that
$$\lb V_1, V_2, V_3 \rb.(\theta^{\alpha_1} \cdots \theta^{\alpha_n})
=0.$$

\noi Performing a similar computation for the space-time
coordinates, we obtain the identities

\beqa \label{deltax} \lb
V_1,V_2,V_3 \rb . x^\alpha &=&
-q^2[\theta,\e_2^\mu][\e_3^\alpha,\e_1{}_\mu]
-q^2[\theta,\e_1^\mu][\e_3^\alpha,\e_2{}_\mu ]\nonumber \\
&&-[\theta,\e_2^\mu][\e_1^\alpha,\e_3{}_\mu]
-[\theta,\e_3^\mu][\e_1^\alpha,\e_2{}_\mu] \\
&&-q[\theta,\e_1^\mu][\e_2^\alpha,\e_3{}_\mu]
-q[\theta,\e_3^\mu][\e_2^\alpha,\e_1{}_\mu] = a^\alpha .\nonumber
 \eeqa

It is important to notice that the $a^\beta$  are complex
numbers. This means that the ``coloration'' of the algebra
\eqref{3poin}, coming from our adapted Jordan-Wigner
transformation gives rise to the algebra \eqref{col3}, which is
manifestly a complex algebra since the structure constants are
complex.  This deserves some explanation. The $\e$ are real
parafermions, therefore the transformation properties
\eqref{trans} ensure that $\delta x$ and $\delta \theta$ are both
real. However, since $a^\beta$ is complex, this means that the
cubic algebra \eqref{3poin} is {\it realised in a
complexification} of the superspace $(x,\theta)$. In other words,
the algebra cannot be realised on a real vector space. This is the
best possible result in this direction using this ansatz.

\section{Conclusion and perspectives}
We have shown that two different cubic algebras (the cubic
extension of the Poincar\'e algebra \eqref{3poin} and the order
two parafermions) ``unify'' in the sense that the latter become
the relevant variables for the construction of an adapted ternary
superspace for the former. In particular this means that we were
able to construct a differential realisation of the algebras
\eqref{3poin}. Having such differential operators, the next step
would be to define some appropriate superfields (depending on
$x^\mu$ and $\theta^\mu$)  and to define certain
 operators which could be interpreted as a
covariant derivative. This possibilities were analysed in
\cite{tern} (together with the study of specific quaternary
extensions of the Poincar\'e algebra). This construction opens the
possibility, using the standard techniques, for the proposal of
interesting physical model constructions based on cubic (and in
general higher order) extensions of the Poincar\'e algebra. One
step to be carefully analyzed under this perspective is the
explicit construction of Lagrangians and other invariant
quantities that provide the experimental confirmation of the model
and fixes to which extent the considered parameters and variables
can be actually identified with well known physical observables.
Further work in this direction is currently in progress.

%%%%%%%%%%%%%%%%%%%%%%%%%%%%%%%%%%%%%%%%%%%%%%%%%%%%%%%%%%%%%
% Doing references:                                         %
%%%%%%%%%%%%%%%%%%%%%%%%%%%%%%%%%%%%%%%%%%%%%%%%%%%%%%%%%%%%%


\begin{thebibliography}{10}

%%%%%%%%%%%%%%%%%%%%%%%%%%%%%%%%%%%%%%%%%%%%%%%%%%%%%%%%%%%%%
%                                                           %
% Command to used is:-                                      %
%                                                           %
%  \bibitem{REFERENCE_LABEL} AUTHORS NAMES,                 %
%  {\it JOURNAL'S NAMES}{\bf VOLUME NUMBER}, PAGE (YEAR).   %
%                                                           %
%  See example below.                                       %
%                                                           %
%%%%%%%%%%%%%%%%%%%%%%%%%%%%%%%%%%%%%%%%%%%%%%%%%%%%%%%%%%%%%

\bibitem{blg}
J.~Bagger and N.~Lambert,
 {\it  Phys.\ Rev.\ }  {\bf D 75}, 045020  (2007)
  [arXiv:hep-th/0611108];
  %%CITATION = PHRVA,D75,045020;%%
 A.~Gustavsson,
 {\it   Nucl.\ Phys.\ }  {\bf B 811}, 66 (2009)
  [arXiv:0709.1260 [hep-th]];
  %%CITATION = NUPHA,B811,66;%%
J.~Bagger and N.~Lambert,
{\it   Phys.\ Rev.\ }  {\bf  D 77}, 065008 (2008)
  [arXiv:0711.0955 [hep-th]].
  %%CITATION = PHRVA,D77,065008;%%
\bibitem{mfm}
 P.~de Medeiros, J.~M.~Figueroa-O'Farrill and E.~Mendez-Escobar,
{\it   JHEP } {\bf 0808}, 045 (2008),
  [arXiv:0806.3242 [hep-th]];
  %%CITATION = JHEPA,0808,045;%%
 P.~de Medeiros, J.~Figueroa-O'Farrill, E.~Mendez-Escobar and P.~Ritter,
{\it   JHEP} {\bf 0904}, 037 (2009),
  [arXiv:0902.4674 [hep-th]].
  %%CITATION = JHEPA,0904,037;%%
\bibitem{cubic}
N.~Mohammedi, G.~Moultaka and M.~Rausch de Traubenberg, {\it Int.\ J.\ Mod.\
Phys.\ } {\bf A19}, 5585  (2004)  [arXiv:hep-th/0305172];
%%CITATION = HEP-TH 0305172;%%
%
G.~Moultaka, M.~Rausch de Traubenberg and A.~Tanasa, {\it  Int.\ J.\ Mod.\
Phys.\ }  {\bf A20}, 5779  (2005)  [arXiv:hep-th/0411198].
%%CITATION = HEP-TH 0411198;%%
%
%
\bibitem{flie}
M.~Rausch de Traubenberg, M.~J.~Slupinski, {\it J.\ Math.\ Phys.\ }  {
\bf 41}, 4556 (2000)  [arXiv:hep-th/9904126];
%%CITATION = HEP-TH 9904126;%%
 M.~Rausch de Traubenberg, M.~J.~Slupinski,
{\it J.\ Math.\ Phys.\ }  {\bf 43}, 5145 (2002) [arXiv:hep-th/0205113].
%%CITATION = HEP-TH 0205113;%%
%
%
\bibitem{gr}
M.~Goze, M.~Rausch de Traubenberg and A.~Tanasa, {\it J. Math. Phys.}
{\bf 48},  093507 (2007)
  [arXiv:math-ph/0603008].
  %%CITATION = MATH-PH/0603008;%%
%
\bibitem{hopf}
M. Rausch de Traubenberg,  {\it J.\ Phys.\ Conf.\ Ser.\ }  {\bf 128},  012060
(2008) [arXiv:0710.5368 [math-ph]];
%%CITATION = 00462,128,012060;%%
M.~Goze and M. Rausch de Traubenberg,
 {\it J. Math. Phys. } {\bf 50},  063508 (2009)
  [arXiv:0809.4212 [math-ph]].
  %%CITATION = ARXIV:0809.4212;%%
%
\bibitem{color}
R.~Campoamor-Stursberg and M.~Rausch de Traubenberg,
{\it J. of Generalized Lie Theory and Appl.}  {\bf 3}, 113 (2009)
  [arXiv:0811.3076 [math-ph]].
  %%CITATION = ARXIV:0811.3076;%%
%
%
\bibitem{noether}
M.~Rausch de Traubenberg,
{\it Pr. Inst. Mat. Nats. Akad. Nauk Ukr. Mat. Zastos.},
{\bf 50}, Part 1, 2, 3, Natsional. Akad. Nauk Ukraïni, \~Inst. Mat.,
Kiev, 2004, pp. 578 [arXiv:hep-th/0312066].
%%CITATION = HEP-TH/0312066;%%
%
\bibitem{cours}
 M.~Rausch de Traubenberg,
 {\it   J. Phys. Conf. Series} {\bf 175}, 012003 (2009)
[arXiv:0811.1465 [hep-th]].
  %%CITATION = ARXIV:0811.1465;%%
%
\bibitem{roby} N. Roby,  Bull. Sc. Math. \textbf{94}, 49 (1970).
%
\bibitem{green}
H.~S.~Green, {\it Phys.\ Rev.\ } {\bf 90}, 270 (1953);
%%CITATION = PHRVA,90,270;%%
%
\bibitem{para}
 O. W. Greenberg, { \it Phys. Rev. Lett.}
\textbf{13}, 598 (1984);
%
  %%CITATION = PRLTA,13,598;%%
 O. W. Greenberg  and A. M. L. Messiah, { \it Phys. Rev.}
\textbf{B136}, 248  (1964);
%%CITATION = PHRVA,136,B248;%%
\textbf{B138}, 1155 (1964).
 %%CITATION = PHRVA,138,B1155;%%
%
Y.~Ohnuki and S.~Kamefuchi, {\it Quantum Field Theory and
Parastatistics}, {\it  Tokyo, Japan: Univ. Pr. (1982) Berlin,
Germany: Springer ( 1982)
 489p}.
%
\bibitem{ter}
I. Bars   and  M. G\"un\"aydin, { \it  J.\ Math.\ Phys.\ }  {\bf
20}, 1977 (1979);
 %%CITATION = JMAPA,20,1977;%%
%cern
I. Bars   and M.  G\"un\"aydin, { \it   Phys.\ Rev.\   } {\bf D22}, 1403
(1980);
 %%CITATION = PHRVA,D22,1403;%%
%
 T.~D.~Palev,
{\it J.\ Math.\ Phys.\ }  {\bf 23}, 1100 (1982);
%%CITATION = JMAPA,23,1100;%%
%
N.~I.~Stoilova and J.~Van der Jeugt, {\it   J. Phys. Conf. Series}
{\bf 128},  012061 (2008).
  [arXiv:math-ph/0611085].
  %%CITATION = MATH-PH/0611085;%%
%
\bibitem{parag}
%
 J.~Beckers and N.~Debergh,
  {\it Int.\ J.\ Mod.\ Phys.\ }   {\bf A8}, 5041  (1993);
  %%CITATION = IMPAE,A8,5041;%%
%
A.~G.~Nikitin and V.~V.~Tretynyk, { \it J.\ Phys.} {\bf A28}, 1655
(1995);
%%CITATION = JPAGB,A28,1655;%%
%
J. Niederle and A.~G.~Nikitin, { \it J.\ Phys.} {\bf A32}, 5141  (1999).
%
\bibitem{ply}
M.~Plyushchay,   {\it  Int.\ J.\ Mod.\ Phys.\ }, 3679  {\bf A15} (2000)
  [arXiv:hep-th/9903130];
  %%CITATION = IMPAE,A15,3679;%%
 S.~Klishevich and M.~Plyushchay,
 {\it  Mod.\ Phys.\ Lett.\ }  {\bf A14}, 2739 (1999)
  [arXiv:hep-th/9905149];
  %%CITATION = MPLAE,A14,2739;%%
F.~Correa, V.~Jakubsky, L.~M.~Nieto and M.~S.~Plyushchay,
 {\it  Phys.\ Rev.\ Lett.\ }  {\bf 101}, 030403  (2008)
  [arXiv:0801.1671 [hep-th]];
  %%CITATION = PRLTA,101,030403;%%
F.~Correa, V.~Jakubsky and M.~S.~Plyushchay,
 {\it  J.\ Phys.\   }{\bf A41},  485303  (2008)
  [arXiv:0806.1614 [hep-th]].
  %%CITATION = JPAGB,A41,485303;%%
%
%
\bibitem{super}
C.~Ahn, D.~Bernard and A.~ Leclair,
 {\it Nucl. Phys. } {\bf B346}, 409 (1990);
%%CITATION = NUPHA,B346,409;%%
%
S.~ Durand,
 {\it Mod. Phys. Lett }{\bf A8},2323  (1993) [hep-th/9305130];
%%CITATION = HEP-TH 9305130;%%
%
N.~ Fleury and M. Rausch de Traubenberg,
 {\it Mod. Phys. Lett.} {\bf A11},899  (1996)
 [hep-th/9510108];
%%CITATION = HEP-TH 9510108;%%
%
A.~Perez, M. Rausch de Traubenberg and P. Simon,
{\it  Nucl. Phys. } {\bf B482}, 325 (1996) [hep-th/9603149];
%%CITATION = HEP-TH 9603149;%%
%
M. Rausch de Traubenberg and P. Simon,
{\it Nucl. Phys. } {\bf B517}, 485 (1998) [hep-th/9606188].
%%CITATION = NUPHA,B517,485;%%
%
\bibitem{az}
%
J.~A.~ de Azc\'arraga and A.~J.~ Macfarlane,
%{\it Group Theoretical Foundations Of Fractional Supersymmetry},
{\it J. Math. Phys. } {\bf 37}, 1115 (1996) [hep-th/9506177];
%%CITATION = HEP-TH 9506177;%%
%
 R.~S.~Dunne, A.~J.~Macfarlane, J.~A.~de Azc\'arraga and J.~C.~P\'erez Bueno,
  {\it  Phys.\ Lett.\ }  {\bf B 387}, 294 (1996)
  [arXiv:hep-th/9607220];
  %%CITATION = PHLTA,B387,294;%%
%
R.~S.~Dunne, A.~J.~Macfarlane, J.~A.~de Azc\'arraga and
J.~C.~P\'erez Bueno,  {\it   Int. J.  Mod. Phys. } {\bf A12}, 3275 (1997)
[hep-th/9610087].
%%CITATION = HEP-TH 9610087;%%
%
\bibitem{tern}
R.~Campoamor-Stursberg and M.~Rausch de Traubenberg,
  arXiv:0907.2149 [hep-th].
  %%CITATION = ARXIV:0907.2149;%%
\bibitem{coloralg}
V.~Rittenberg and D.~Wyler,
  {\it J.\ Math.\ Phys.\ } {\bf 19}, 2193  (1978);
 %%CITATION = JMAPA,19,2193;%%
%
V.~Rittenberg and D.~Wyler,
  {\it Nucl.\ Phys.\ }  {\bf B139}, 189 (1978);
%
M.~Scheunert,
 {\it  J.\ Math.\ Phys.\ }  {\bf 20}, 712  (1979);
  %%CITATION = JMAPA,20,712;%%
%
H.~S.~Green and P.~D.~Jarvis,
  {\it  J.\ Math.\ Phys.\ }  {\bf 24}, 1681 (1983);
  %%CITATION = JMAPA,24,1681;%%
%
J.~Lukierski and V.~Rittenberg,
  {\it Phys.\ Rev.\ } {\bf D18}, 385 (1978).
  %%CITATION = PHRVA,D18,385;%%
%
\end{thebibliography}
\end{document}